\documentclass[11pt,a4paper]{article}
\pdfoutput=1
\usepackage{jcappub}
\usepackage[T1]{fontenc} 

\usepackage{graphicx}
\usepackage{dcolumn}
\usepackage{bm}
\usepackage{hyperref}
\usepackage{color}
\usepackage{soul}

\RequirePackage[english]{babel}
\usepackage{bm}      
\usepackage{amssymb}   
\usepackage{amsmath}   
\usepackage{graphicx}	
\usepackage{morefloats}
\usepackage{subfigure}
\usepackage{hyperref}
\usepackage{nicefrac}

\usepackage{color}

\begin{document}

\title{Assessing the role of the Kelvin-Helmholtz instability \\ at the QCD cosmological transition}

\author[]{V. R. C. Mour\~{a}o Roque}
\author[]{and}
\author[]{G. Lugones}

\affiliation[]{Universidade Federal do ABC, Centro de Ci\^encias Naturais e Humanas, Avenida dos Estados 5001- Bang\'u, CEP 09210-580, Santo Andr\'e, SP, Brazil.}
\emailAdd{victor.roque@ufabc.edu.br}
\emailAdd{german.lugones@ufabc.edu.br}

\abstract{We performed numerical simulations with the  PLUTO code in order to  analyze the non-linear behavior of the Kelvin-Helmholtz instability in non-magnetized relativistic fluids. The relevance of the instability at the cosmological QCD phase transition was explored using an equation of state based on lattice QCD results with the addition of leptons. The results of the simulations were compared with the theoretical predictions of the linearized theory.
For small Mach numbers up to  $M_s \sim 0.1$ we find that both results are in good agreement. However,  for higher Mach numbers, non-linear effects are significant. In particular, many initial conditions that look stable according to the linear analysis are shown to be unstable according to the full calculation. 
Since according to lattice calculations the cosmological QCD transition is a smooth crossover, violent fluid motions are not expected.  Thus, in order to assess the role of the Kelvin-Helmholtz instability at the QCD epoch, we focus on simulations with low shear velocity and use  monochromatic as well as random perturbations to trigger the instability.  We find that the Kelvin-Helmholtz instability can strongly amplify turbulence in the primordial plasma and as a consequence it may increase the amount of primordial gravitational radiation. Such turbulence may be relevant for the evolution of the Universe at later stages and may have an impact in the  stochastic gravitational wave background.   }


\maketitle

\flushbottom

\section{Introduction}
%
It is well accepted that during its early stages the Universe went through several transitions that changed its composition. In particular, the quark-hadron transition (QCDt), occurred about 10 $\mu$s after the initial big bang and lasted approximately  1 $ \mu$s. In this period, a hot unconfined quark-gluon plasma was converted, as the Universe expanded and cooled, into a confined hadronic phase. 
Most of the early studies about this epoch assumed that a first-order transition occurred, in which hadronic bubbles were nucleated within the quark-gluon plasma when the temperature approached the critical value $T_c \sim 100-200$ MeV. 
This assumption leads to a rich phenomenology involving bubble growth through deflagrations and/or detonations \cite{Witten1984,Cleymans1986,Suonio1986}, formation of seeds of the cosmic magnetic field \cite{Cheng1994}, inhomogeneous nucleation \cite{Ignatius2001}, and production of gravitational waves (GWs) \cite{Hogan1986}, with possible consequences for pre-galactic structure formation and Big Bang nucleosynthesis. 

The recent detection of gravitational waves by Advanced LIGO \cite{Abbott2016} and the expected launch of new detectors within the next years (e.g. eLISA/NGO \cite{Binetruy2012}, Big Bang Observatory (BBO) \cite{Phinney2004} and TOBA \cite{Ishidoshiro2011}), brings new attention to the production of primordial GWs  because they could be the only tool able to provide direct information from cosmological phase transitions.  
The first works on gravitational radiation from the QCDt focused on the GWs generated by the nucleation, growth and collision of hadronic bubbles (see e.g. \cite{ Turner1990,Miller1995,Kosowsky1993, Kamionkowski1994}). Later, the attention was directed to the stochastic background of GWs arising from a period of turbulence that the collisions produced \cite{Kosowsky2002,Nicolis2004,Kahniashvili2005, Kahniashvili2008,Caprini2006, Kahniashvili2010a,Grojean2007}. 

More recently, lattice QCD  calculations introduced a new key ingredient in the analysis, showing that the QCDt at low chemical potentials and high temperatures (which is the condition expected in the early Universe) is merely a crossover \cite{Aoki2006,Borsanyi2014}. This indicates that the transition at the QCD epoch occurred very smoothly, inhibiting the generation of relics that could be observed nowadays.  Nonetheless, even in this adverse scenario, it has been shown  that if the fluid contains velocity fluctuations produced in previous epochs with $(\Delta v) /c  \gtrsim 10^{-2}$ and/or temperature fluctuations with $\Delta T/T_c \gtrsim  10^{-3}$,  the QCDt would produce a gravitational wave signal above the eLISA's sensitivity curve \cite{Amaro2012} at frequencies larger than $\sim 10^{-4}$ Hz \cite{Mourao2013}. 
These lower bounds were obtained from one-dimensional relativistic hydrodynamic simulations using a realistic lattice QCD equation of state and were extended to three dimensions through an appropriate averaging procedure that takes advantage of some symmetries of the problem \cite{Mourao2013}. 
However, being based on one-dimensional simulations, such analysis cannot take into account some features (naturally arising in two and three dimensional studies) that could increase the turbulence in the fluid and consequently the production of gravitational waves. 

An interesting phenomenon that occurs in severals systems and at different scales is the Kelvin-Helmholtz instability (KHI). It is one of the most important hydrodynamical instabilities and plays a significant role in various parts of astrophysics. The instability occurs when a velocity shear is present within a continuous fluid or across fluid boundaries. The shear is converted into vorticity that, subject to secondary instabilities, cascades generating turbulence \cite{McNally}. The relatively simple initial condition necessary to trigger it, which does not require gravity or magnetic fields, makes it very likely to have occurred in the Universe during the QCDt. 
The first studies about the KHI date from the late nineteenth century and were pioneered by Helmholtz and Kelvin \cite{Helmholtz,Kelvin} considering incompressible fluids. 
More recently, many studies, most of them numerical, considered the KHI in the relativistic limit and its consequences for astrophysical objects \cite{Turland,Ferrari,Bodo:2004id,Bucci,Beckwith,Berne,Radice}. In particular, Bodo et al. \cite{Bodo:2004id} reexamined the KHI for relativistic flows studying the linear stability of an interface separating two non-magnetized relativistic fluids in relative motion. They inferred that, in an appropriate reference frame, it is possible to find analytic solutions to the dispersion relation. 

In this paper, we focus on the KHI in the specific context of the primordial Universe at the QCDt. 
Using a lattice QCD equation of state, we perform two-dimensional relativistic hydrodynamic simulations with the numerical code PLUTO  \cite{Mignone2007}\footnote{http://plutocode.ph.unito.it} testing the limits of the linear theory presented in Bodo et al. \cite{Bodo:2004id} and the implications of non-linear effects. 
We analyze the linear growth and the subsequent non-linear saturation of the instability for different values of the Mach number, estimate the amount of turbulence injected in the fluid, and explore some clues about the role of the KHI in the generation of primordial GWs.

This article is organized as follows. In Sec. \ref{sec:basic_equations} we present the relativistic hydrodynamic equations, the lattice QCD equation of state, the linear theory for the relativistic KHI, the initial conditions for the hydrodynamic simulations, and some details concerning the numerical methods. In Sec. \ref{sec:results} we describe our results, focusing on the growth and saturation of the KHI and its role in the QCDt. In Sec. \ref{sec:conclu} we present our conclusions.

\section{Basic equations}
\label{sec:basic_equations}

\subsection{Relativistic hydrodynamics}

To investigate properly the dynamics and evolution of the primordial fluid, we should solve the equations of hydrodynamics in the context of general relativity. However, if we focus on a sufficiently small portion of the Universe during a small enough period of time, we can consider a flat metric and use the hydrodynamic equations in the special relativistic limit without taking into consideration the cosmic expansion. In fact, the Hubble time at the QCD transition, $ t_{QCD} \sim 10^{-5} $ s \cite{Vega2006}, is much longer than the  characteristic time $\tau_c$ and the saturation time $\tau_s$ of the KHI which are in the range $10^{-7} - 10^{-9}$ s as will be verified later.

Another property of the system that should be considered is the viscosity of the fluid. According to experiments in heavy ions colliders \cite{RHIC,Song2011,Heinz2005} the thermalized quark-gluon plasma created in collisions is a strongly coupled plasma which behaves like an almost ideal fluid, with viscosity per entropy density in the range  $1 < 4 \pi (\eta / s)_{QGP} < 2.5$ \cite{Song2011,Heinz2012}, i.e. approaching the Kovtun-Policastro-Son-Starinets lower bound  $\eta / s \geq 1 / (4 \pi)$ \cite{Kovtun2005}. 
Theoretical estimates of the quark-gluon plasma viscosity in the early Universe \cite{Ahonen1999} agree qualitatively with the experimental values obtained at the  Relativistic Heavy Ion Collider (RHIC) and the Large Hadron Collider (LHC), in spite of the comparatively different thermodynamic conditions. Although the viscosity may be increased significantly if electrons, muons, photons and (specially) neutrinos are included in the quark-gluon plasma,  it is still very small and the Reynolds number is very large.

Based on the above considerations, we shall use special relativistic non-dissipative hydrodynamics to describe a portion of the primordial Universe at the QCD epoch.  Adopting $c=1$, the equations read \cite{Weinberg}: 
\begin{eqnarray}
\dfrac{\partial (Wn_B)}{\partial t} + \nabla\cdot (Wn_B \mathbf{v}) = 0 ,  \label{eq:relat_form1}\\ 
\dfrac{\partial v^i}{\partial t} + v^i \cdot\nabla\mathbf{v} + \dfrac{1}{W^2(p + E)}\left( \dfrac{\partial p}{\partial x^i} + v^i\dfrac{\partial p}{\partial t} \right)  = 0  , \label{eq:relat_form2}\\ 
W\dfrac{\partial p}{\partial t} + W\mathbf{v} \cdot\nabla p - \dfrac{\partial[W(p + E)]}{\partial t} - \nabla\cdot(W (p + E) \mathbf{v}) = 0, \label{eq:relat_form3}
\end{eqnarray}
where $n_B$ is the net baryon number density, $v$ is the fluid velocity,  $E$ is the total energy density, $p$ is the pressure,  and $W$ denotes the Lorentz factor,
\begin{equation*}
	W = \dfrac{1}{\sqrt{1 - \mathbf{v}^2}}.
\end{equation*}

At the QCD epoch, the net baryon number density is very close to zero. Thus,  Eq. \eqref{eq:relat_form1} can be neglected and the evolution of the system over short time intervals is controlled only by Eqs. \eqref{eq:relat_form2} and \eqref{eq:relat_form3}. 
In order to complete the above set of equations, it is needed an equation of state (EOS) of the form
\begin{equation}
	p = p(E),
\end{equation}
from which the speed of sound can be evaluated according to 
\begin{equation}\label{eq:sound_def}
	c_s^2 \equiv \left. \dfrac{\partial p}{\partial E} \right\rvert_s 
\end{equation}
where $s$ is the entropy per particle.

In order to solve the fluid equations we use the PLUTO code, which is a freely distributed modular code for the solution of nonlinear systems of conservative partial differential equations of the mixed hyperbolic/parabolic type under different regimes (e.g., classical/relativistic fluid dynamics, Euler/MHD). The implementation is based on the well-established framework of Godunov type, shock- capturing schemes where an upwind strategy (usually a Riemann solver) is employed to compute fluxes at zone faces \cite{Mignone2012}. It also has a friendly interface and is well documented.

The PLUTO code solves the hydrodynamic equations in a form that involves the proper rest mass density $\rho$, the internal energy $\epsilon$ and the pressure $p$. Some of these quantities, e.g.  $\rho$, are not well defined in the early Universe since the primordial soup at the QCD epoch consists mainly of effectively massless degrees of freedom. However,  a numerically useful correspondence between the variables at the PLUTO code and Eqs. \eqref{eq:relat_form1}$-$\eqref{eq:relat_form3} can be established because the mathematical form of both sets of equations is very similar. In practice, adopting the correspondence $\rho \leftrightarrow n_B$ with $\rho$ being a very small number, $E \leftrightarrow \rho + \rho \epsilon \approx \rho \epsilon$ and $p \leftrightarrow p$ we can solve Eqs. \eqref{eq:relat_form1}$-$\eqref{eq:relat_form3} using the PLUTO code. Since  $\rho$ is numerically very small, the quantity $\rho \epsilon$ in the PLUTO code can be identified with our total energy density $E$. We have tested different small values of $\rho$ consistent with vanishing baryon number density $n_B$ and checked that the results of the simulations are the same.

\subsection{Equation of state}
%
\begin{figure}[tb]
    \centering
\includegraphics[width = 0.65 \textwidth]{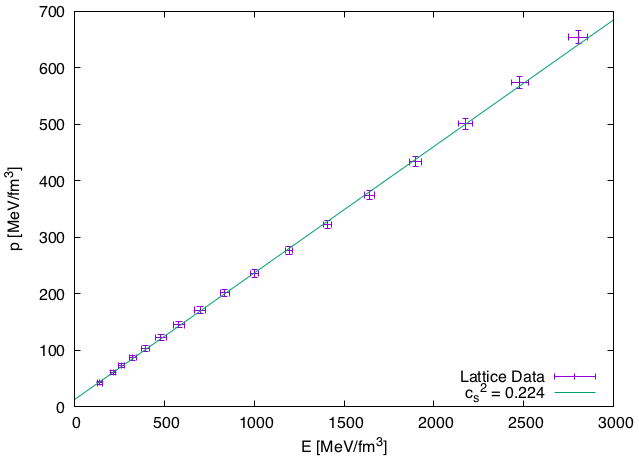}
\caption{Pressure $p$ versus total energy density $E$ describing a crossover transition obtained using data from lattice QCD calculations \cite{Borsanyi2010} plus the contribution of leptons and photons in the temperature range from $120$ to $200$ MeV. The solid line represents the linear fit of these data with an ideal EOS with $\Gamma = 1.224$ and $c_s = 0.473$.}
\label{fig:EOS}
\end{figure}

The EOS has to describe the thermodynamic behavior of the primordial fluid during all the transition that happens at temperatures about $\sim 150 - 200 $ MeV. In this temperature range, we consider that the existing particles are up, down and strange quarks, gluons, neutrinos ($\nu_\tau, \nu_\mu, \nu_e$), muons, electrons, photons and their antiparticles. To construct the EOS, we considered the results of lattice QCD simulations {at vanishing baryon chemical potential} obtained by the Wuppertal-Budapest collaboration (see Ref.\ \cite{Borsanyi2010}), with $N_f = 2 + 1$, i.e. two light (up and down) and one heavy (strange) quarks and added the contribution of a gas of non-interacting neutrinos ($\nu_\tau, \nu_\mu, \nu_e$), muons, electrons, photons and their antiparticles, all them described by an EOS of the form $E^{\prime} = p^{\prime}/3 = g \pi^2 T^4 / 90$ with $g = 7/8 \times 14 + 2 = 14.25$. 

In order to set up an EOS compatible with the PLUTO code, we restrict the data to the temperature range from $120$ to $200$ MeV, which results  in a quite linear relationship between the total energy density and the pressure, as can be seen in Fig.\ \ref{fig:EOS}. Then, we fit the data with an ideal EOS of the form: 
\begin{equation}\label{eq:idealEOS}
	p = E (\Gamma - 1),
\end{equation}
where the ratio of specific heats is $ \Gamma = 1.224 $, and the sound speed is a constant with value $ c_s = 0.473 $. In the raw data, the sound speed is not constant but varies smoothly in the range $ 0.46 - 0.52 $, showing a few percent variation with respect to the linear fit that doesn't change the results significantly.

\subsection{Linearized theory: dispersion relation and growth rate of the Kelvin-Helmholtz instability}
\label{linear_theory}

In this section we give a brief resume of the analytic results obtained by Bodo \textit{et al.} \cite{Bodo:2004id} about the stability of an interface separating two nonmagnetized relativistic fluids in relative motion. The desired analytical dispersion relation can be obtained by finding the perturbative solutions of the linearized version of the relativistic equations given in Eqs. \eqref{eq:relat_form1}-\eqref{eq:relat_form3}.

The simple initial condition to start the instability is
\begin{equation}\label{eq:khi_init_rel}
\mathbf{v} = \left\{ 
\begin{array}{rc}
(U,0,0) & \mbox{if}\quad y > 0, \\
(-U,0,0) & \mbox{if}\quad y < 0,
\end{array}\right.
\end{equation}
where $ U $ is positive, the two fluids are initially in pressure equilibrium, and have the same proper density $ \rho $. In the rest frame, indicated here by a tilde, a generic three-dimensional perturbation of the flow variables $\delta\tilde{q} $ (here $q$ is one of  $\rho$, $\mathbf{v}$, $p$) has the form
\begin{equation}
\label{eq:pert_rest}
\delta \tilde{q}_{\pm} \propto \exp [i(\tilde{k}_{\pm} \tilde{x} + \tilde{l}_{\pm}\tilde{y} + \tilde{m}_{\pm} \tilde{z} - \tilde{\omega}_{\pm} \tilde{t}) ].
\end{equation}

Here the  subscripts $ \pm $ refer to the fluids that are initially at the regions $ y \gtrless 0 $,  and $ (\tilde{k}, \tilde{l}, \tilde{m}) $ are the spatial wave vectors related to the frequency $ \omega $ by the dispersion relation 
\begin{equation}\label{eq:freq_rest}
\tilde{\omega}^2_{\pm} = (\tilde{k}^2_{\pm} + \tilde{l}^2_{\pm} + \tilde{m}^2_{\pm})c_s^2.
\end{equation}

In our simulations we shall use the laboratory frame, where the fluids have the initial condition given by Eq.\ \eqref{eq:khi_init_rel}. In this frame the solutions will still have the form
\begin{equation}\label{eq:pert_lab}
	\delta q_{\pm} \propto \exp [i(kx + l_{\pm}y + mz - \omega t) ].
\end{equation}

The dispersion relation in this frame can be obtained by a Lorentz transformation of Eq.\ \eqref{eq:freq_rest}, 
\begin{equation}
\label{eq:freq_lab}
W^2(\omega\mp kU)^2 = \left[W^2\left(k\mp\omega\dfrac{U}{c^2}\right)^2 + l^2_\pm + m^2 \right]c^2_s.
\end{equation}
Introducing the growth rate $ \phi = \nicefrac{\omega}{c_sk} $, the classical Mach number $ M = U/c_s$, the relativistic Mach number $ \mathcal{M} = \nicefrac{WM}{W_s} $, and the parameters $ W_s  = (1 - \nicefrac{c_s^2}{c^2})^{1/2}$ and $\beta= U/c$, it is possible to write Eq. \eqref{eq:freq_lab} in the form of a fifth-order polynomial
\begin{equation}\label{eq:rel_disp}
\phi \left[ \left(\dfrac{\phi}{M}\right)^{4} (\mathcal{M}^2 + 2\beta^2) - 2 \left(\dfrac{\phi}{M}\right)^2 (\mathcal{M}^2 + 1 + \alpha^2 - \beta^2) +  (\mathcal{M}^2 - 2 - 2\alpha^2)   \right] = 0.
\end{equation}

The parameter $ \alpha $ is related to the angle $ \theta $ between the velocity of the fluid and the wave number in the $ x-y $ plane by
\begin{equation*}
\cos\theta = \dfrac{1}{\sqrt{1+\alpha^{2}}}.
\end{equation*}

\begin{figure}[t]
    \centering
\includegraphics[angle=0, width = 0.60 \textwidth]{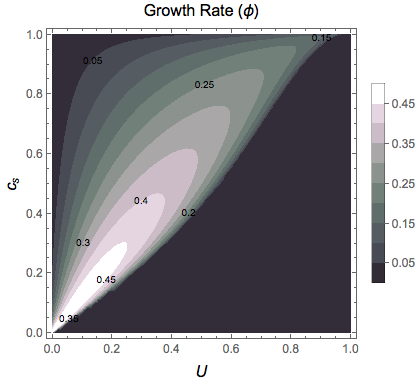}
\caption{Contours with different values of the growth rate $\phi$ as a function of the sound speed $c_s$ and the shear velocity $U$. Within the darkest region the fluid is stable. Within the closed region with different gray shades the KHI may grow according to the linear analysis. }
\label{fig:phi} 
\end{figure}

\begin{figure}[t]
    \centering
	\includegraphics[angle=0, width = 0.65 \textwidth]{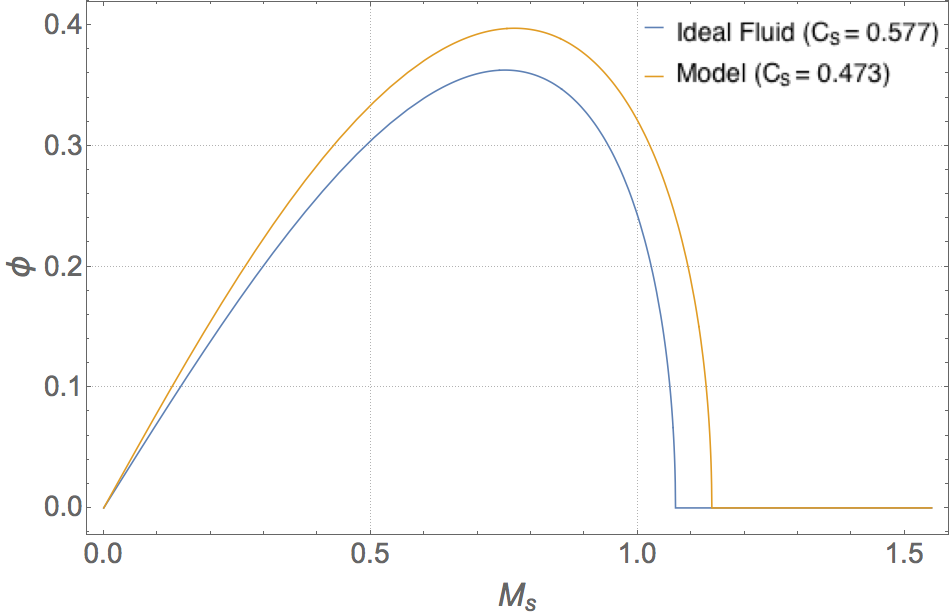}
	\caption{The growth rate $\phi$ as a function of the Mach number for two different EOS with constant sound speed: the ultra relativistic fluid and the lattice QCD EOS shown in Fig. \ref{fig:EOS}.}
	\label{fig:phi_comp}
\end{figure}

In this work we will focus on the specific case in which the perturbation is parallel to the flow and consequently $ \cos\theta = 1 $. Thus, the solutions of Eq.\ \eqref{eq:rel_disp} read \cite{Bodo:2004id}:
\begin{subequations}\label{eq:rel_sol}
	\begin{gather}
	\phi = 0 \label{eq:phi0}, \\[0.4cm]
	\dfrac{\phi^{2}}{M^{2}} = \dfrac{\mathcal{M}^{2} + 1 - \beta^2 \pm \sqrt{4\mathcal{M}^{2}(1-\beta^2) + (1 + \beta^2)^2} }{\mathcal{M}^2 + 2\beta^2}. \label{eq:phi1}
	\end{gather}
\end{subequations}
The solution with $\phi =0$ and the solution of Eq. \eqref{eq:phi1} with $+$ sign are trivial and correspond to stable modes. 
Instabilities are found in some ranges of Eq.\eqref{eq:phi1} with the negative sign. In Fig. \ref{fig:phi} we show the contours with different values of the growth rate as a function of the sound speed and the shear velocity. The darkest area corresponds to a stable condition. Within the unstable region the  clearer the area the higher the growth rate. Notice that the sound speed of the fluid,  which is given by Eq.\ \eqref{eq:sound_def}, determines the range of shear velocities that  generate  instabilities. 
As stated before, the lattice EOS can be well fitted by a fluid with constant sound speed $ c_s = 0.473 $, which is below the value for the usual ultra relativistic ideal plasma ($ c_s = \nicefrac{1}{\sqrt{3}}\approx 0.577 $). Therefore, the range of Mach numbers that generate unstable conditions is slightly wider than for the ideal fluid, as showed in Fig.\ \ref{fig:phi_comp}.

\subsection{Initial conditions and numerical methods}\label{sec:IC}

The initial condition used in the simulations is based on Ref.\ \cite{Beckwith}, but expressing the shear velocity as a function of the Mach number of the shear, $ M_s $, in order to  distinguish between subsonic and supersonic initial flows. The initial velocity in  the $ x- $direction is given  by: 
\begin{equation}
\label{eq:khMsx}
\mathrm{v^x}(y) = \left\{ \begin{array}{cc}
c_s M_{s}\tanh\left(\frac{y-0.5}{a}\right) & \mathrm{if}\quad y > 0, \\ 
[0.25cm]  - c_s M_{s}\tanh\left(\frac{y+0.5}{a}\right) & \mathrm{if}\quad y \leq 0,
\end{array}
\right. 
\end{equation}
where $a = \nicefrac{\mathrm{L}}{100}$ and $ \mathrm{L} $ is the size of computational domain in $ x- $direction. The parameter $ a $ was chosen to avoid the growth of perturbations in the same scale as the computational cells \cite{Bucci}, which occur when an interface is too sharp. 
We used several values of $ M_s $, ranging from $ M_s = 0.05 $ to $ 2.1 $, i.e. we explored the response of the fluid well beyond the unstable region   shown in Fig.\ \ref{fig:phi_comp}. 
This range is sufficient for obtaining data that can be compared with the analytic results, thus improving the understanding of the consequences of the nonlinear effects neglected in the linearized theory presented in Sec. \ref{linear_theory}.

The instability is seeded by applying a monochromatic perturbation on the transverse velocity of the form
\begin{equation}\label{eq:khMsy}
\mathrm{v^y}(x,y) = \left\{ \begin{array}{cl}
\mathrm{A}_0 c_s M_{s} \sin(k_x x)\exp\left[\frac{-(y-0.5)^2}{\sigma}\right] &\mathrm{if}\, y > 0, \\[0.25cm] -\mathrm{A}_0 c_s M_{s} \sin(k_x x)\exp\left[\frac{-(y+0.5)^2}{\sigma}\right] & \mathrm{if}\, y \leq 0.
\end{array}
\right. 
\end{equation}
Here, $ k_x = \nicefrac{2\pi}{\mathrm{L}} $ was chosen with the largest possible length, $ \sigma = \nicefrac{\mathrm{L}}{10} $ describes the characteristic length within which the amplitude of the perturbation decays by a factor $e$, and we adopt a typical value $A_0 = 10^{-2}$. 
The pressure is homogeneous in all the domain with value $ 276.49\,{\mbox{ MeV}}/{\mbox{fm}^{-3}} $, corresponding to a temperature of 170 MeV.

The computational domain is a box with sides $L$ and $2L$, where $L= 1$m was chosen because it would be the maximum size of a disturbance generated in a preceding primordial phase transition, specifically at the electroweak epoch \cite{Mourao2013}. To evaluate the convergence of the results, we performed  simulations with two different resolutions: low $ 512 \times 1024 $ (L) and high $ 1024 \times 2048 $ (H).

The PLUTO code was configured to perform WENO3 reconstruction of primitive variables. To compute the inter-cell fluxes a HLLC Riemann solver has been employed (as recommended in \cite{Beckwith,Mignone2009}), while third order accuracy in time has been achieved using a Runge-Kutta scheme.

\section{Results}
\label{sec:results}

\subsection{Growth rates and saturation times}

The time evolution of the fluid perturbations begins with a period of linear growth which is followed by a stage of saturation presenting some fluctuations but without any meaningful growth. 
For the complete understanding of the phenomenon, it is relevant to determine the growth rate $ \phi $, the saturation time $ \tau_s $ and the interrelation between them. From the definition of $ \phi $ it is possible to obtain a characteristic time $ \tau_c = \lambda/(c_s\phi) $ where $ \lambda $ is the perturbation length, set as $ \lambda = L = 1 $m. 
%
\begin{figure}[tb]
    \centering
	\includegraphics[angle=0, width = 0.65 \textwidth]{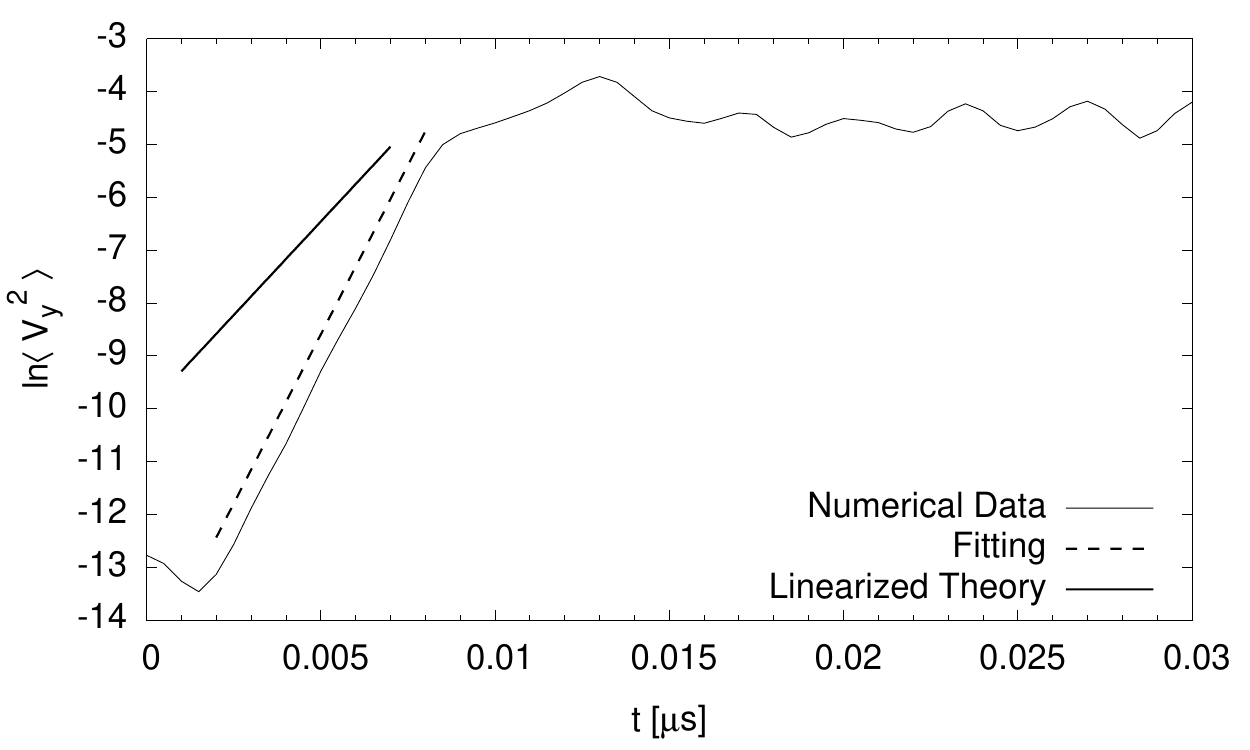}
	\caption{Time evolution of the logarithm of the mean square of velocity perturbation $ v_y $ (solid thin line). This data is used to determine the growth rate $ \phi $ and the saturation time $ \tau_s $ of the simulation. The growth rate is determined by the linear fit (dashed line) in this plot to the time interval of the linear growth. The solid black line indicates the growth rate predicted by linear theory, shown in Eq.\eqref{eq:phi1}.} 
	\label{fig:v2}
\end{figure}

\begin{figure}[tb]
    \centering
	\includegraphics[angle=0, width = 0.55 \textwidth]{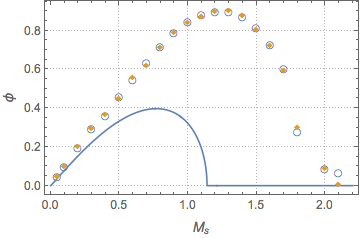}
	\caption{Growth rate versus shear Mach number. The solid line indicates the analytic result of the linearized theory while the circles and filled diamonds show the simulated data with low resolution $(512\times 1024)$ and high resolution $(1024\times2048)$ respectively. Note that the numerical results have already converged to the solution, not requiring higher resolutions.}
	\label{fig:phi_theor_simul}
\end{figure}

The numerical growth rate can be obtained from the time evolution of the mean square of the transverse velocity. As shown in Eq.\ \eqref{eq:pert_rest}, the perturbation develops exponentially, with $ \langle v_y^2 \rangle \propto \exp[{2t}/{\tau_c}] $ giving a straight line in a logarithmic plot. Such behavior can be seen in the initial evolution of the fluid shown in Fig.\ \ref{fig:v2}. The growth rate can be determined from the angular coefficient of this line obtained by a linear regression of the data. A similar method was employed in Ref.\ \cite{Keppens1999}. The saturation time $\tau_s$ was estimated by taking the first local maximum after the growth period. 

In Fig.\ \ref{fig:v2}, we show the mean square of the transverse velocity for a simulation with $ M_s = 0.8 $ and compare with the linear regression of the data and the analytic prediction. The simulation shows a growth rate significantly larger than the linearized theory. 
A similar result is obtained in simulations for most values of $M_s$ as can be seen in Fig.\ \ref{fig:phi_theor_simul}. Notice that non-linear effects stretch the curve without changing the shape qualitatively: at low $M_s$ there is a small positive slope, followed by a maximum around $ M_s = 1.2 - 1.3 $ and ending with a larger negative slope at $M_s \sim 2$, near stabilization. The two used resolutions converged into a unique solution, not requiring the use of a thinner grid.

A quantitative comparison between analytic and numerical results can be seen in Table \ref{tab:data}. As the Mach number grows, the difference between them is larger. Even stable conditions predicted by the linearized theory ($M_s \gtrsim 1.2$) became highly unstable in the simulations. 
The numerical values obtained for $ \tau_c $ and $ \tau_s $ are roughly of the same order, showing that the KHI saturates typically in one characteristic time. Since the analytic $ \tau_c $ is always larger than the numerical  $ \tau_c $ and $ \tau_s $, it serves as a rough upper limit of what can be expected from a full calculation.

\begin{table}[!htb]
	\centering
	\begin{tabular}{c|c|c|c|c|c|}
		\cline{2-6}
		& \multicolumn{2}{c|}{Theoretical Results} & \multicolumn{3}{c|}{Numerical Results} \\ \hline
		\multicolumn{1}{|c||}{$M_s$} & $\phi$            & $\tau_c \, [\mu s]$        & $\phi$   & $\tau_c \, [\mu s]$   & $\tau_s  \, [\mu s]$  \\ \hline \hline
		\multicolumn{1}{|c||}{0.05}  & \; 0.040 \;        & 0.177           & \; 0.047  \;  & \; 0.150 \;      & \; 0.130 \;     \\ \hline
		\multicolumn{1}{|c||}{0.1}   & 0.079         & 0.090           & 0.094    & 0.075      & 0.055     \\ \hline
		\multicolumn{1}{|c||}{0.2}   & 0.155          & 0.046           & 0.196    & 0.036      & 0.032     \\ \hline
		\multicolumn{1}{|c||}{0.3}   & 0.224          & 0.031           & 0.292    & 0.024      & 0.022     \\ \hline
		\multicolumn{1}{|c||}{0.4}   & 0.284          & 0.025           & 0.361    & 0.020      & 0.017     \\ \hline
		\multicolumn{1}{|c||}{0.5}   & 0.334          & 0.021           & 0.456    & 0.015      & 0.013     \\ \hline
		\multicolumn{1}{|c||}{0.6}   & 0.371          & 0.019           & 0.545    & 0.013      & 0.011     \\ \hline
		\multicolumn{1}{|c||}{0.7}   & 0.393           & 0.018           & 0.634    & 0.011      & 0.010      \\ \hline
		\multicolumn{1}{|c||}{0.8}   & 0.396          & 0.018           & 0.717    & 0.010      & 0.009     \\ \hline
		\multicolumn{1}{|c||}{0.9}   & 0.376          & 0.019           & 0.788    & 0.009      & 0.009     \\ \hline
		\multicolumn{1}{|c||}{1}     & 0.320          & 0.022           & 0.845    & 0.008      & 0.009     \\ \hline
		\multicolumn{1}{|c||}{1.1}   & 0.187          & 0.038           & 0.883    & 0.008      & 0.008     \\ \hline
		\multicolumn{1}{|c||}{1.2}   & -                 & -               & 0.895    & 0.008      & 0.008     \\ \hline
		\multicolumn{1}{|c||}{1.3}   & -                 & -               & 0.895    & 0.008      & 0.008     \\ \hline
		\multicolumn{1}{|c||}{1.4}   & -                 & -               & 0.869    & 0.008      & 0.008     \\ \hline
		\multicolumn{1}{|c||}{1.5}   & -                 & -               & 0.813    & 0.009      & 0.009     \\ \hline
		\multicolumn{1}{|c||}{1.6}   & -                 & -               & 0.727    & 0.010      & 0.008     \\ \hline
		\multicolumn{1}{|c||}{1.7}   & -                 & -               & 0.602    & 0.012      & 0.015     \\ \hline
		\multicolumn{1}{|c||}{1.8}   & -                 & -               & 0.277    & 0.025      & 0.019     \\ \hline
		\multicolumn{1}{|c||}{2}     & -                 & -               & 0.087    & 0.081      & 0.080      \\ \hline
		\multicolumn{1}{|c||}{2.1}   & -                 & -               & 0.065    & 0.108      & 0.200       \\ \hline
	\end{tabular}
    	\caption{Comparison between the theoretical prediction of the linearized theory and the  results of the numerical simulations. }
	\label{tab:data}
\end{table}

\subsection{Kelvin-Helmholtz instability at the QCD epoch}

Since the primordial QCDt is most probably a crossover and not a first order phase transition, high Mach numbers associated with violent phenomena seem unlikely.
For this reason, we focus mainly on the simulations with small $M_s$. Specifically, we use a shear velocity that, according to Ref. \cite{Mourao2013}, is the lowest one that could be detected by the eLISA GW observatory, i.e. $ M_s \sim 0.05 $ \cite{Mourao2013}.

\begin{figure}[tb]
\centering
\includegraphics[width = 0.70 \textwidth]{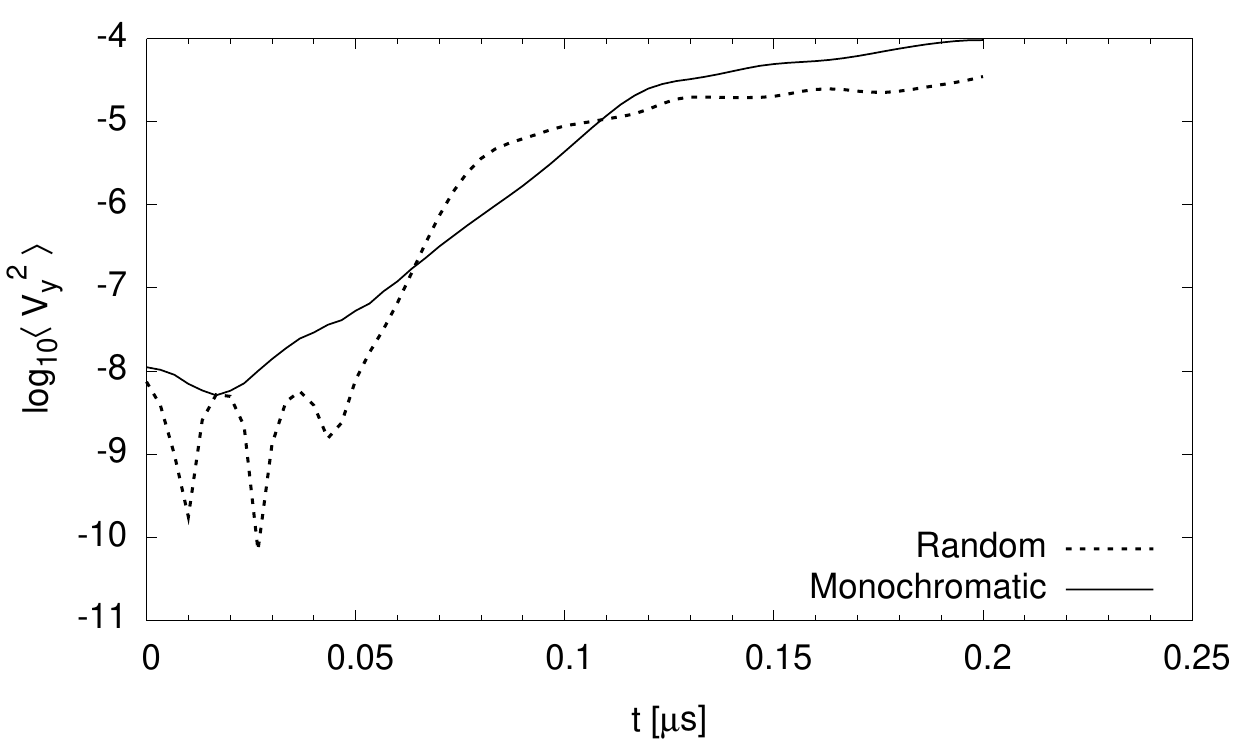}
\caption{Time evolution of the logarithm of the mean square of velocity perturbation for random and monochromatic initial conditions. }
\label{fig:v2_00_comp}
\end{figure}

In addition to the monochromatic perturbations analyzed in the previous section, we also simulate  initial conditions with random perturbations, changing the sine in Eq. \eqref{eq:khMsy} by a random number from $0$ to $1$ in order to represent a more plausible scenario. A comparison of the mean square velocity of these simulations is shown in Fig. \ref{fig:v2_00_comp} for a case with $ M_s = 0.05 $. 
Using the same methods explained above, we find that the simulation with a random perturbation has a growth rate $ \phi_{random} = 0.131$ which is significantly larger than the value $ \phi_{mono} = 0.047 $ for the monochromatic perturbation (see Table \ref{tab:data}). However, as apparent from Fig. \ref{fig:v2_00_comp}, saturation is attained roughly at the same time in both cases. This is mostly due to the fact that  the simulation with a random perturbation has a transient time lasting approximately $0.045\mu$s before the beginning of the linear growth while the monochromatic growth starts much earlier, at around $ 0.015\mu$s. 

As discussed in Ref. \cite{Mourao2013}, a peak in the gravitational wave spectrum can be expected as a consequence of an exponential variation in the fluid velocity which in the case with $M=0.05$ happens on a timescale of $\tau_{mono} \approx 0.115 \mu$s and $\tau_{random} \approx 0.045 \mu$s, as seen in Fig. \ref{fig:v2_00_comp}. Since gravitational wave emission is associated with the level of turbulent motion in the fluid, we expect a maximum in the spectrum at a frequencies  $ \tau_{mono}^{-1} \sim 8.7 \times 10^6$ Hz and  $\tau_{random}^{-1} \sim 2.2 \times 10^7$ Hz. These waves are redshifted  in their way to the present Universe, thus the present day frequency is given by \cite{Maggiore1999}:
\begin{eqnarray}
f_0 &=& 8\times 10^{-14} f_* \left( \frac{100}{g_*}\right)^{1/3}  \left( \frac{1 \mbox{GeV}}{T_*}\right)  \mathrm{Hz} \label{eq:freq_today}, 
\end{eqnarray}
where the subscript $0$ corresponds to present values and the subscript  $*$ to the values at the QCD epoch. Using $T_* \approx 170$ MeV and $g_* = 43/4$, peaks would be found now at $ f_{random} \sim 2.2 \times 10^{-5} $Hz for the random perturbation and at $ f_{mono} \sim 0.86 \times 10^{-5} $Hz for the monochromatic one, both close to the sensitivity window of eLISA \cite{Amaro2012}.

Although the random perturbation has a small saturation time, it was effective in triggering several secondary instabilities that cascade generating turbulence. In fact, at the end of simulations with both initial conditions the transverse velocities are increased by $\sim 2$ orders of magnitude and the typical final transverse velocity becomes of the order of the initial shear velocity $U$. This stresses the potential of the KHI to generate turbulence in the primordial plasma.

\begin{figure*}
\centering
\includegraphics[width = .227\textwidth]{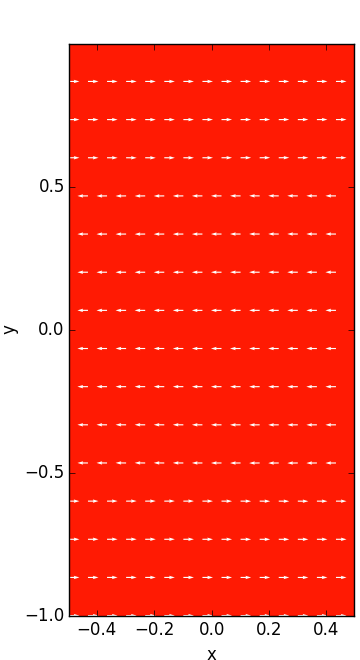}
\includegraphics[width = .185\textwidth]{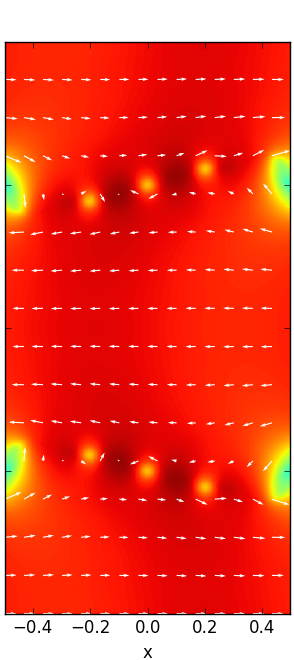}
\includegraphics[width = .185\textwidth]{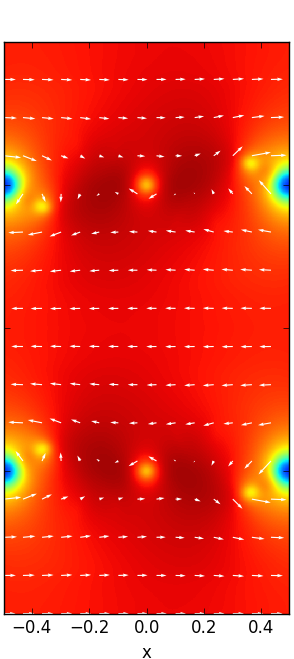}
\includegraphics[width = .185\textwidth]{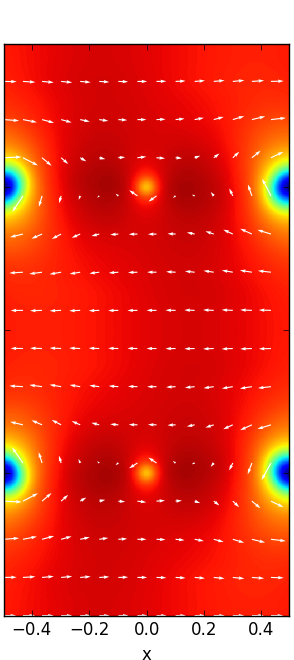}
\includegraphics[width = .185\textwidth]{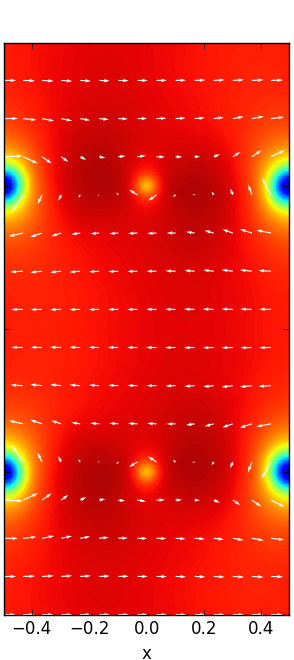}  \\
\includegraphics[width = 0.65\textwidth]{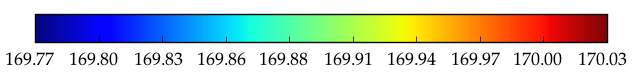}
\caption{Evolution of the temperature (color scale in MeV) and the fluid velocity (white arrows) for a  monochromatic perturbation. From left to right we have $t = 0.000,\, 0.125,\, 0.250,\, 0.375 $ and $ 0.500 \mu$s. For $t > 0.250 \mu$s, there is little variation of the temperature and velocity profiles over the time. At early times, big eddies are generated at the positions $(\pm 0.5, \pm 0.5)$, and stay there until the end. Four of the smaller eddies are absorbed by the larger ones and the remaining two stay fixed at $(0.0,\pm 0.5)$. Notice that, due to the periodic boundary conditions, the four observed eddies are actually only two.}
\label{fig:prs_mono}
\end{figure*}

\begin{figure*}
\centering
\includegraphics[width = .227\linewidth]{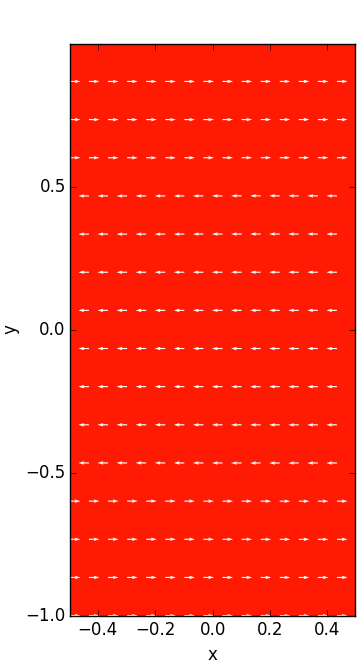}
\includegraphics[width = .185\textwidth]{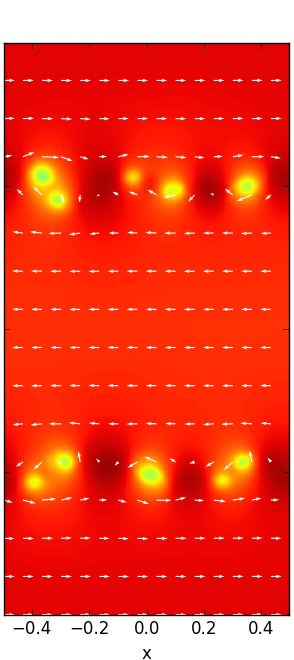}
\includegraphics[width = .185\textwidth]{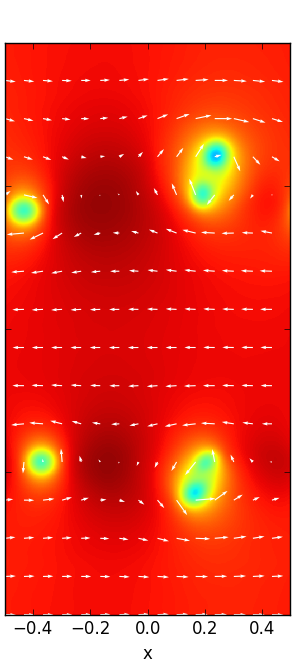}
\includegraphics[width = .185\textwidth]{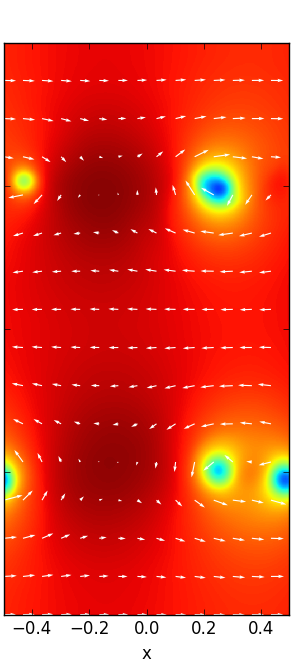}
\includegraphics[width = .185\textwidth]{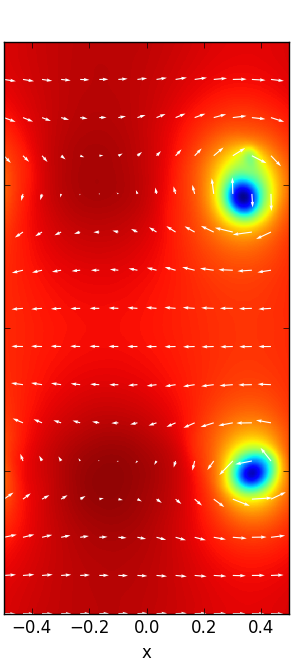}  \\
\includegraphics[width = 0.65\textwidth]{figures/bar_temp.png}
\caption{Same as in the previous figure but for random perturbations. From left to right we have $ t = 0.000, 0.125, 0.250, 0.375 $ and $ 0.500 \mu$s. The time evolution of the fluid is more pronounced than in  Fig. \ref{fig:prs_mono}. Initially, several small eddies are formed which merge together forming the two large eddies seen in the rightmost snapshot. In spite of the differences, the simulations with random and monochromatic perturbations end with very similar temperature and velocity profiles with eddies located at the shear of the fluxes $(y = \pm 0.5)$.}
\label{fig:prs_rdm}
\end{figure*}

Notice that the KHI is able to produce and maintain temperature inhomogeneities of one part in $10^3$ (see Figs. \ref{fig:prs_mono} and \ref{fig:prs_rdm}). An important aspect of these perturbations is that they involve predominantly the velocities rather than the density or the temperature. As a consequence, in spite of its small scale, this turbulence may be relevant for  further stages of the Universe. 
This is so because at the recombination epoch the effective speed of sound of the cosmic fluid drops by some orders of magnitude and  the  turbulent motions (previously subsonic) would become supersonic following the decoupling time \cite{Goldman1993}. Therefore, part of the primordial turbulent eddies could dissipate into shocks and generate large density contrasts that may have an impact in the post decoupling epoch.

\newpage
\section{Summary and Conclusions}
\label{sec:conclu}

In this work we studied non-linear effects of the Kelvin-Helmholtz instability in non-magnetized relativistic fluids and explored their relevance  for the cosmological QCD  transition. To this end, we performed hydrodynamic simulations with the numerical code PLUTO using an EoS based on lattice QCD results.

We first reviewed the analytic results  obtained in Ref. \cite{Bodo:2004id}  for the growth rate of the instability focusing on the specific case of an initial perturbation parallel to the shear velocity. In Fig. \ref{fig:phi} we show all the possible initial configurations of velocities and sound speeds that generate unstable conditions according  the linearized theory. Then, we compare the numerical growth rate obtained from our simulations with the analytic results.
As shown in Table \ref{tab:data} and in Fig.\ \ref{fig:phi_theor_simul}, the linearized theory agrees with the numerical growth rate only in the case of very small Mach numbers, up to  $M_s \sim 0.1$. For higher Mach numbers, non-linear effects are significant; in fact, many initial conditions that look stable according to the linear analysis are shown to be unstable according to the full calculation. 

Since the QCD transition is a crossover, we don't expect a violent behavior of the fluid and small Mach numbers are more likely. Thus, in order to analyze the influence of the KHI in the dynamic evolution of the cosmic fluid, we adopt two different initial conditions with low shear velocity but different instability trigger: one with a monochromatic perturbation (as in the linearized theory) and another with random perturbations (which appears as more realistic in the cosmological context).  In both cases, the simulations show that perturbations in the velocity are increased by $\sim 2$ orders of magnitude, showing that the presence of the KHI can amplify turbulence in the primordial plasma and as a consequence it may increase the amount of primordial gravitational radiation. 

In Figs. \ref{fig:prs_mono} and \ref{fig:prs_rdm} we show the time evolution of  the temperature and the  velocity obtained from the simulations. Despite the differences in the initial conditions and in the early stages, the final state of the fluid is quite similar in both cases. Small eddies arise in the fluid involving small density and temperature variations but with velocities of the same order of the initial shear.  If produced at the QCD epoch, this turbulence may be relevant for  later stages of the Universe in spite of its  very small scale. In fact, some effects could be expected at the recombination epoch when the effective speed of sound of the cosmic fluid drops by some orders of magnitude. Previously subsonic  turbulent motions may become supersonic following the decoupling time \cite{Goldman1993} generating shock waves, energy dissipation and large density contrasts.  
Finally, since turbulence at the QCD epoch may reprocess part of the spectrum of gravitational waves produced at inflation, further studies are worthwhile because they may help to optimally extract inflationary information from the stochastic gravitational wave background.

\acknowledgments
G.L. acknowledges the Brazilian agencies Conselho Nacional de Desenvolvimento Cient\'{\i}fico e Tecnol\'ogico (CNPq) and Funda\c c\~ao de Amparo \`a Pesquisa do Estado de S\~ao Paulo (FAPESP) for financial support.

\end{document}